\documentclass[reprint,
 amsmath,amssymb,
 aps,
 prc
]{revtex4-2}

\usepackage{graphicx}
\usepackage{dcolumn}
\usepackage{bm}
\usepackage{epstopdf}
\usepackage[dvipsnames]{xcolor}

\begin{document}

\preprint{APS/123-QED}

\title{$A_{\beta}[E_{\beta}]$ in $^{37}$K decay: new physics with opposite $\beta$ helicity}

\author{Melissa Anholm$^{1,2}$}
\author{J.A. Behr$^{1,2,5}$}
\author{D.G. Melconian$^{3,4}$}
\author{G. Gwinner$^{2}$}%
\author{A. Gorelov$^1$}
\author{J.C. McNeil$^{5,1}$}
\author{B. Fenker$^{3,4}$}
\author{S. Behling $^{3,4}$}
\email{corresponding author: behr@triumf.ca}
\affiliation{
$^1$TRIUMF, 4004 Wesbrook Mall, Vancouver, BC V6T 2A3 Canada\\
$^2$ University of Manitoba, Department of Physics and Astronomy, MB Canada\\
$^3$ Cyclotron Institute, Texas A\&M University, 3366 TAMU, College Station, Texas 77843-3366, USA\\
$^4$ Department of Physics and Astronomy, Texas A\&M University,
  4242 TAMU, College Station, Texas 77843-4242, USA\\
$^5$ University of British Columbia, Department of Physics and Astronomy, 6224 Agricultural Road, Vancouver, B.C. V6T 1Z1 Canada
}




\date{\today}

\begin{abstract}
  By extending our analysis and simulations of our $^{37}$K $\beta$-decay data set to allow the $\beta$ asymmetry with respect to nuclear spin to vary with $\beta$ energy $E_{\beta}$, we have gained sensitivity to new physics that depends on a helicity factor for the $\beta$, $m_\beta/E_\beta$.
  In particular, we constrain Lorentz scalar and tensor quark-lepton interaction strengths at a sensitivity complementary to the similar Fierz interference term in neutron $\beta$ decay.
  Our result for that new physics is a Fierz interference term
 $b$ = 0.033 $\pm$ 0.084 (stat) $\pm$ 0.039 (syst),
  consistent with the standard model electroweak interaction value $b=0$. We consider presently achieved complementarity to  $\beta$-decay and particle physics experiments, along with projectable technical improvements to our method.
\end{abstract}

\maketitle


\section{\label{sec:level1}Introduction} 

The standard model electroweak interaction always makes left-handed helicity leptons paired with right-handed helicity antileptons. 
In a previous publication~\cite{Fenker2018}, we constrained new physics coupling to flipped helicity for both the $\beta$ and the $\nu$, i.e. quark-lepton interactions Lorentz transforming like vector (V) + axial vector (A) currents, rather than the weak interaction's V-A Lorentz current structure.
Our observable,
the asymmetry of $\beta$ emission with respect to nuclear spin A$_{\beta}$,
was predicted to be independent of total $\beta$ energy E$_{\beta}$ either for V-A or V+A.

Here we extend simulations and analysis techniques to allow for different new physics that couples a $\nu$ with weak interaction helicity to a $\beta^+$ with the same sign of helicity as the $\nu$, which multiplies the result by a helicity factor $m_\beta/E_\beta$ needed to produce that $\beta$ with opposite helicity.
Having noted in Ref.~\cite{Fenker2018} that one dominant systematic, $\beta$ scattering off surfaces, strongly depends on $E_\beta$ as it approaches $m_{\beta}$, we find that the energy dependence  complicates the simulations and analysis considerably.

\section{Theory and Methods}

New physics that couples outgoing leptons and antileptons with same helicity-- rather than opposite helicity as in the standard model weak interaction-- produces a decay contribution dependent on a helicity factor of the massive $\beta$, $m_{\beta}/E_{\beta}$, the Fierz interference term between the new physics and the standard model matrix elements. 

The decay rate expression then adds the Fierz term $b$: 

\begin{equation}
\frac{dW}{dE} = F(E,Z) p E p_\nu E_\nu (1 + b m/E +\hat{I} \cdot (A_\beta \frac{\vec{p_\beta}}{E})),
\end{equation}

\noindent
The normalized coefficient of $\cos{\theta}$, where $\theta$ is the angle between the $\beta$ and the initial nuclear polarization, becomes a function of the $\beta$ energy:

\begin{equation}
  A_\beta[E] = A_\beta/(1+b m/E) \approx A_\beta  (1- b m/E)
  \end{equation}

\noindent We write the approximate expression to emphasize that sensitivity to $b$ in any given decay scales with $A_\beta$. The value of $A_\beta$ in $^{37}$K is reasonably large compared to other well-measured cases like the neutron and $^{19}$Ne, making $^{37}$K decay more sensitive to $b$ by that simple figure of merit.

In the Lee-Yang Lagrangian, $b$ is produced by a Lorentz scalar interaction interfering with the standard model Lorentz vector interaction, and a Lorentz tensor interaction interfering with a Lorentz axial vector interaction~\cite{Jackson1957}:

\begin{equation}
  b = \frac{\pm \gamma [ (C_S+C_S') + \rho^2 (C_T+C_T')]}{1 + \rho^2 + (C_S^2+C_S'^2+\rho^2 (C_T^2+C_T'^2))/2}
\end{equation}

\noindent where $\gamma$=$\sqrt{1-Z^2 \alpha^2}$. For isobaric mirror decay

  \begin{eqnarray}
    A_\beta = \frac{\frac{\pm \rho^2}{J+1}(C_TC_T'-1)}{1 + \rho^2 + (C_S^2+C_S'^2+\rho^2(C_T^2+C_T'^2))/2} \nonumber\\
      + \frac{\rho \sqrt{\frac{J}{J+1}} (C_SC_T'+C_S'C_T -2)}{1 + \rho^2 + (C_S^2+C_S'^2+\rho^2(C_T^2+C_T'^2))/2}    
  \end{eqnarray}

  \noindent where $\rho = \frac{g_A}{g_V}\frac{M_{GT}}{M_F}$
  is
  determined by us below from our collaboration's measured $^{37}$K $fT$ value~\cite{Shidling2014},
  together with an average $M_F$ from $0^+ \rightarrow 0^+$ decays~\cite{TownerHardy2020}.
   The $\pm$ signs are for $\beta^\mp$ decay.



We note that the Lee-Yang Lagrangian assumes zero-range interactions, i.e. interactions carried by bosons with mass much greater than the $\beta$ decay energy release, so that terms from expanding the boson propagator are negligible. It otherwise characterizes completely general types of Lorentz structures of quark-lepton interactions. 

There is new physics less universal than the Lee-Yang Lagrangian.
When quarks are combined into nucleons by QCD, there are two induced lepton-nucleon currents that flip sign when n$\leftrightarrow$p, termed 2nd-class currents. The lepton-nucleon vector current acquires what is termed the induced scalar $g_S k_\mu$, where $k_\mu$ is the momentum transfer, which divergence does not vanish. The form factor $g_S$ is nonzero in the standard model for strangeness-changing baryon $\beta$ decay~\cite{Holstein1982}, but vanishes in agreement with the conserved vector current hypothesis for first-generation $\beta$ decay~\cite{Holstein1982,Tsushima1988}. 
This 2nd-class induced scalar contribution to the lepton-nucleon vector current
introduces the same $E_\beta$ dependence to observables as the $b$ term~\cite{Ormand1989,Szybisz1981,Holstein1984}.
The term can also be written
as $\frac{e}{A}\frac{m_\beta}{m_{\rm nucleon}} \frac{m_\beta}{E_\beta}$,
where A is the nucleus mass number, and $e/A$ is fit to data~\cite{Holstein1984}.
Although in the impulse approximation $e$ scales with $A$~\cite{Holstein1974}, a similar scaling expected both in impulse approximation and in meson models for a second-class induced tensor form factor~\cite{Holstein1974,Lipkin1971} was found to be incorrect in more self-consistent complete models~\cite{Kubodera1973}.
For hypothetical models with a constant form factor $e$, effects would then naturally fall with nuclear mass as well as average $E_\beta$.

A 2nd-class term in the axial vector lepton-nucleon current, termed the induced tensor term $\frac{g_T}{m_{\rm nucleon}}\sigma_{\mu \nu}k_\nu$, has different dependence on $E_{\beta}$ than the electron-quark Lorentz tensor interaction~\cite{Holstein1974,Wilkinson2000}. It can be generated by non-standard-model meson exchange and nucleon-nucleon contact interactions~\cite{Wilkinson2000,Kubodera1973,Kubodera1977}, and there are tight constraints from tailored nuclear $\beta$ decay observables, in particular the $\beta$-alignment correlation which vanishes in Gamow-Teller decay~\cite{Minamisono2011}. In this parameterization of 2nd-class tensor interactions, one term does contribute equally to neutron and nuclear $\beta$ decay, while two other complicated nuclear matrix elements can vary with any nucleus, only one of which vanishes in isobaric mirror decays~\cite{Kubodera1973,Kubodera1977}. Relating constraints to those on $C_T$ is nontrivial, so we don't comment further below.

A number of experimental methods worldwide measure small changes in the $\beta$ energy spectrum itself. Instead, we focus on the asymmetry of emission of $\beta$'s with respect to the initial spin. Since the asymmetry is not changing much with energy-- mainly with $v/c$-- this minimizes dependence on systematics like the $\beta$ detector lineshape response.

%
%
%
%

\subsection{Recoil-order corrections}

We treat in our standard model description
Coulomb and 1st and 2nd-order recoil corrections
according to Holstein~\cite{Holstein1974}.
Several recoil-order terms in isobaric mirror decay are given by the conserved vector current hypothesis-- e.g. weak magnetism and 2nd-order in recoil electric quadrupole moment analog are given by their respective nuclear moment differences in parent and progeny, while the first-class induced tensor term vanishes.
The $\mu$ moment of $^{37}$K, and thus the weak magnetism term, are small, while the $\beta$ energy endpoint near 5 MeV makes the 2nd-order corrections similar in magnitude.
We found that corrections can be combined to add
$\approx$ -0.0028 $E_\beta$/$E_0$ to $A_\beta$~\cite{Fenker2018}.
A number of careful corrections to Ref.~\cite{Holstein1974} 
were shown small in our case~\cite{HayenYoung2019}.


\section{Experiment }

\subsection{\label{sec:level2}TRIUMF Neutral Atom Trap}

\begin{figure}[htb]
    \centering
    \includegraphics[width=0.70\linewidth]{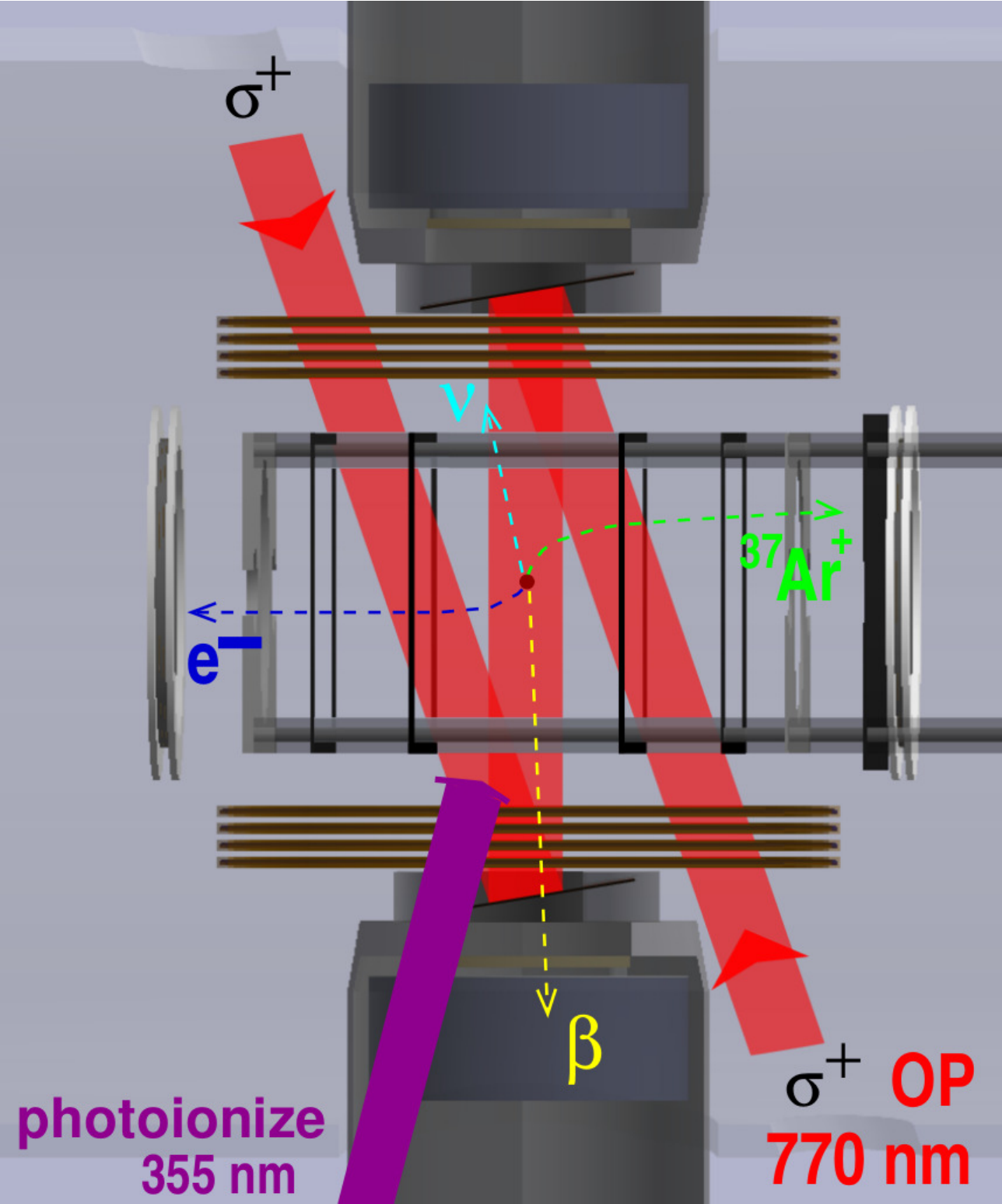}
    \caption{TRINAT during the optical pumping time. Shown are $\beta$ telescopes (double-sided silicon strip detector DSSD $\Delta E$ backed by plastic scintillator), mirrors for optical pumping light and its beams, magnetic field coils, electric field electrodes, and microchannel plates for electron and ion detection. Distance between trap cloud and ion MCP is 9.7 cm.}
    \label{fig:trap}
\end{figure}

Fig.~\ref{fig:trap} is a side view of
the detection apparatus of  TRIUMF's Neutral Atom Trap for $\beta$ decay (TRINAT). Not shown is the collection trap from a vapor cell cube, nor the push beams between traps~\cite{Swanson1998}.

Using 8x10$^{7}$/s mass-separated $^{37}$K delivered from the TRIUMF/ISAC isotope separator on-line facility, we trapped on average 10,000 $^{37}$K atoms during the data-taking time. ISAC, by using a microstructured TiC target, has recently demonstrated 4 times more yield.

\subsection{Polarization by optical pumping}

We refer the reader to our {\em in situ} measurement of the nuclear polarization by atomic methods, Ref.~\cite{FenkerNJP2016} and the Supplementary Material of Ref.~\cite{Fenker2018}.

During the polarization time, we switch the magnetic field from the trap's quadrupole to a uniform 2 Gauss field pointed up, and apply circularly polarized light along the quantization axis.
      Once we start the optical cycle, atoms increase spin to maximum, then stop
      absorbing the S$_{1/2}$ to P$_{1/2}$ light. 

      By atomic physics probes of this process independent of the $\beta$ decays, we measured nuclear polarization $I_z/I = 0.991 \pm 0.001$, which contributes negligibly to our uncertainty budget.

\subsection{Geometry and Detectors}

An electric field collects
$^{37}$Ar ions produced in $^{37}$K $\beta^+$ decay,
along with
photoions from the 355 nm laser in Fig.~1, 
to an MCP with 78 mm active diameter located 9.7 cm away.
By comparing trajectories with a detailed finite element calculation, we found our simulations could assume a uniform 150 V/cm field. The photoion information helps determine the cloud polarization, center in space, and spatial distribution. The $^{37}$Ar ion info is not used in this paper's analysis.

Decay by $\beta^+$ naturally feeds the negative Ar$^{-1}$ ion. We constrained contributions from the known Ar$^{-1}$ $\tau$=250 ns metastable state by the lack of a significant tail on the timing spectrum between shakeoff electron and $\beta$: nevertheless, to eliminate possible bias on $A_\beta$ from loss of detection efficiency from the delayed shakeoff emission and its dependence on the recoil energy and direction,  
we don't impose a vertical position cut on the shakeoff electron detector~\cite{Fenker2018}.

We used 0.30 mm thick
double-sided silicon strip detectors as Ref.~\cite{Fenker2018}, 
requiring both X and Y strips above energy threshold and similar calibrated energy deposited.
Our $E$ detectors are plastic 3.5x9 cm scintillators shown, characterized further in Ref.~\cite{Ozen2023}.

\subsection{Data and interpretation}

\begin{figure}[htb]
    \centering
    \includegraphics[width=\linewidth]{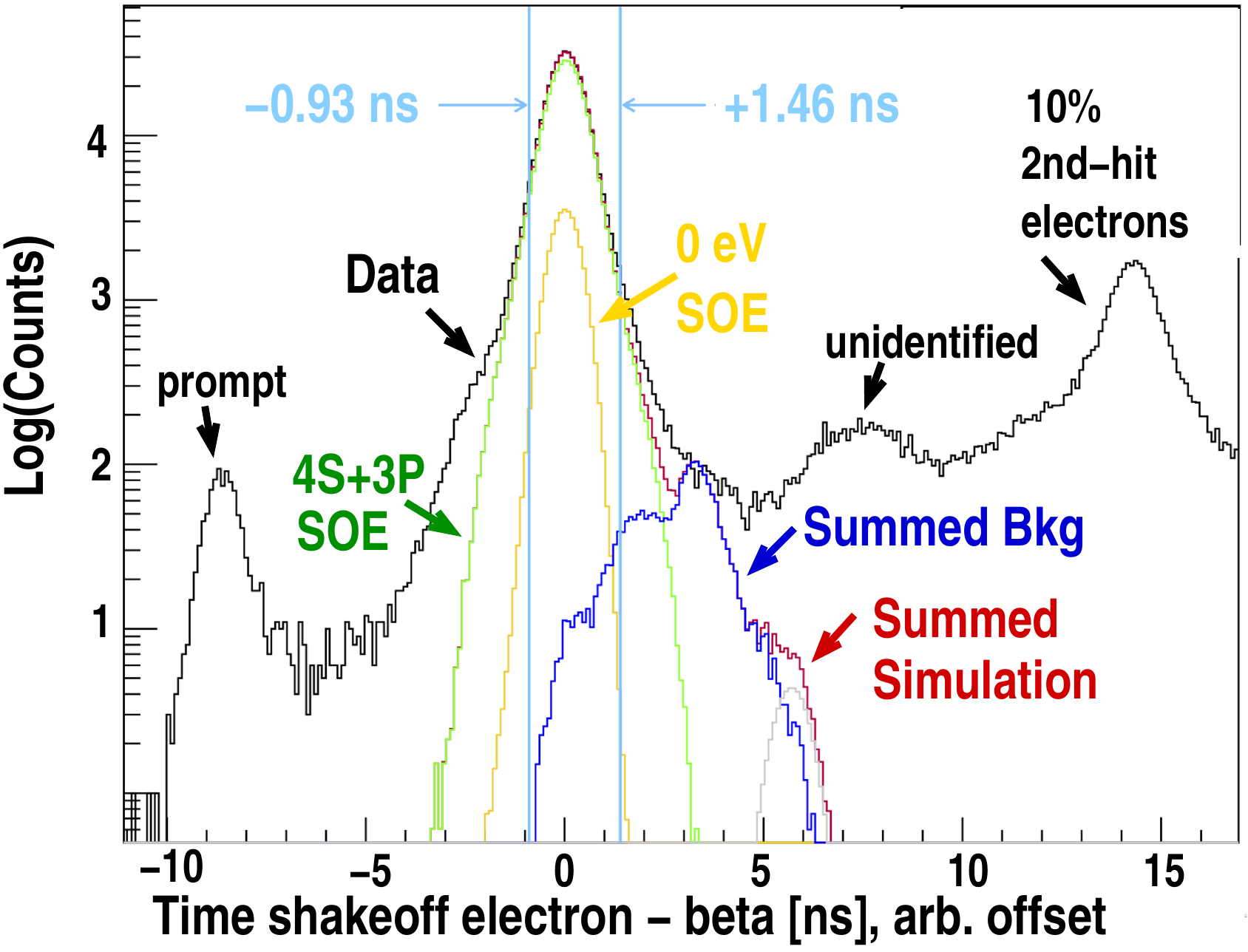}
    \caption{Improved shakeoff electron (SOE) timing wrt the $\beta$ compared to Fig.~4 of Ref.~\cite{Fenker2018}, and the resulting smaller background. The data timing cut is shown. The summed simulation reproduces the data well in the region of interest. 10\% of real coincidences are from electrons not firing eMCP, scattering, and returning in the E field 14 ns later-- note this late 10\%
of the      prompt $\beta$-$\gamma$ coincidences do not fall in the ROI.}
      \label{fig:timing}
\end{figure}

We found it important in Ref.~\cite{Fenker2018} to remove backgrounds in our geometry from decays of untrapped atoms by requiring the $\beta^+$ to be in coincidence with shakeoff atomic electrons (SOE's).
To improve our knowledge of the asymmetry at low $E_{\beta}$ to study $b$,
we found the relative timing to be even more important.
So we removed a residual timing walk in the scintillator,
produced at low $E_\beta$ pulse height, which had not been 
removed by the constant fraction discriminator.
The resulting improved timing (Fig.~\ref{fig:timing}) nearly removes a small background from untrapped atoms on surfaces that is more apparent at low $E_\beta$.
The simulated background is scaled to a tail of the timing spectrum,
as we do not know what fraction of untrapped
atoms sticks long enough to decay. This background was one of our larger systematics in Ref.~\cite{Fenker2018}, while with the improved timing the correction from the simulation shown is nearly negligible and introduces small uncertainty.  Also, we can now understand and model well the measured timing signal from the known cloud size using a combination of 9\% very low-energy electrons-- probably Ar$^-$ autoionization-- 77\% 4S electron shakeoff characterized by their 4.5 eV binding energy, and 14\% 3P electron shakeoff characterized by their 18 eV binding energy. The SOE energy spectra use the sudden approximation calculations with hydrogenic wavefunctions by Levinger~\cite{Levinger1953}. 

\subsubsection{Simulations}

\begin{figure}[htb]
    \centering
    \includegraphics[width=1.00\linewidth]{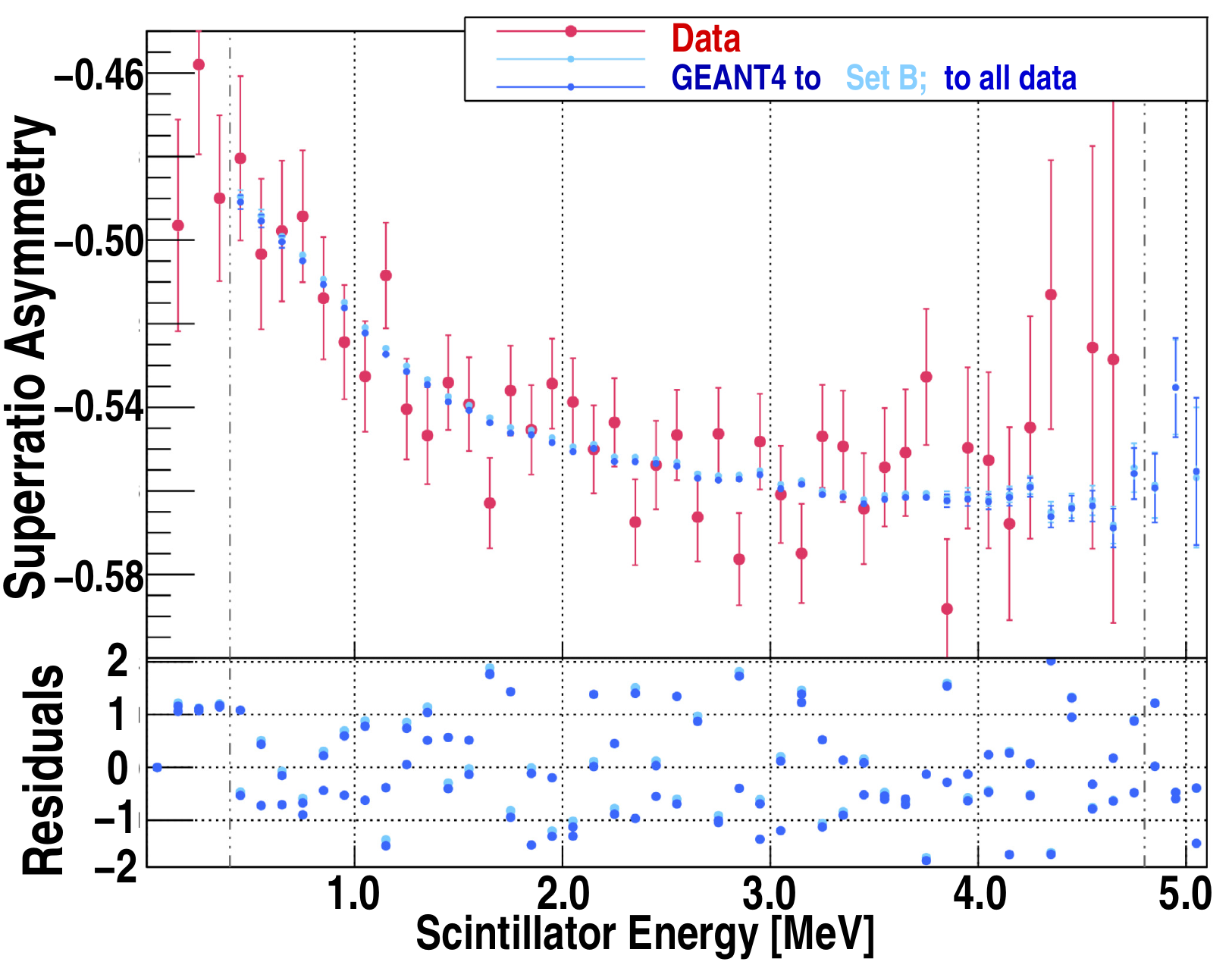}
    \caption{Superratio asymmetry from ``Dataset B'' (one-half of the useful data) with 1$\sigma$ statistical uncertainties, compared to the GEANT4 simulation with best fit $b$ and $A_\beta$ to Set B and to all data (almost same), and the residuals.}
        \label{fig:fit}
\end{figure}

We show our GEANT4 simulation against half the data in Fig.~\ref{fig:fit}. A superratio observable combining up and down $\beta$ detectors and up and down nuclear spin polarization
measures
the $\beta$ asymmetry.
The inflection region between 1.5 and 2 MeV is critical, and the improved fit here removes a small systematic apparent in our previous result Ref.~\cite{Fenker2018}. The fit threshold is the same as in Ref.~\cite{Fenker2018}, determined by  the Compton edge from 511 keV annihilation radiation in our plastic scintillator $E$ detector, together with energy loss in our optical pumping mirrors and our DSSD $\Delta E$ detector. The other half of the data has a slightly different energy calibration, so we do not choose to show a combined plot.

\subsection{Extraction of $A_{\beta}[E_\beta]$}

We extract $A_{\beta}$ and $b$ from this info. We calculate $\chi^2$ from the difference between the superratio of all data and simulations with varied $A_{\beta}$ and $b$.
To extract our two degrees of freedom, we then look for this total $\chi^2$ to increase from its minimum by 2.3 when varying $A_\beta$ and $b$~\cite{PDG2024}. 
Allowing $b$ and $A_\beta$ to vary independently produces our main result,
$b$ = 0.033 $\pm$ 0.084 (stat) $\pm$ 0.0039 (syst) and $A_{\beta} = -0.5738 \pm 0.0082 {\rm (stat)} \pm 0.0041 {\rm (syst)}$.
The linear correlation between $A_\beta$ and $b$ reflected in Fig.~4 can be expressed by $A_\beta = -0.5707-0.090*b$.

We find more sensitivity to new physics by adding information.
We also show the standard model prediction for $A_{\beta}=-0.5706$ in Fig.~4, given $\rho$ and its uncertainty from the measured $^{37}$K $Ft$ of our collaboration~\cite{Shidling2014}, and taking $M_F$ from the measured $0^+ \rightarrow 0^+$ average $Ft_0$ value~\cite{TownerHardy2020}. In Fig.~4 we are not considering the variation in $Ft_0$ with $C_S+C'_S$,
as we wish to show $^{37}$K complementarity to such observables below.
We further show $\chi^2$ contours as black ellipses in Fig.~4 including the contribution from the experimental uncertainty on $^{37}$K $Ft$. 
When Lorentz scalar and tensor interactions are considered below, we correlate $b$ variations with changes in $A_{\beta}$ by Eqs.~3-4, and plot the resulting combined constraint on those couplings in Fig.~5.




\begin{figure}[htb]
    \centering
   \includegraphics[width=0.7\linewidth,angle=90]{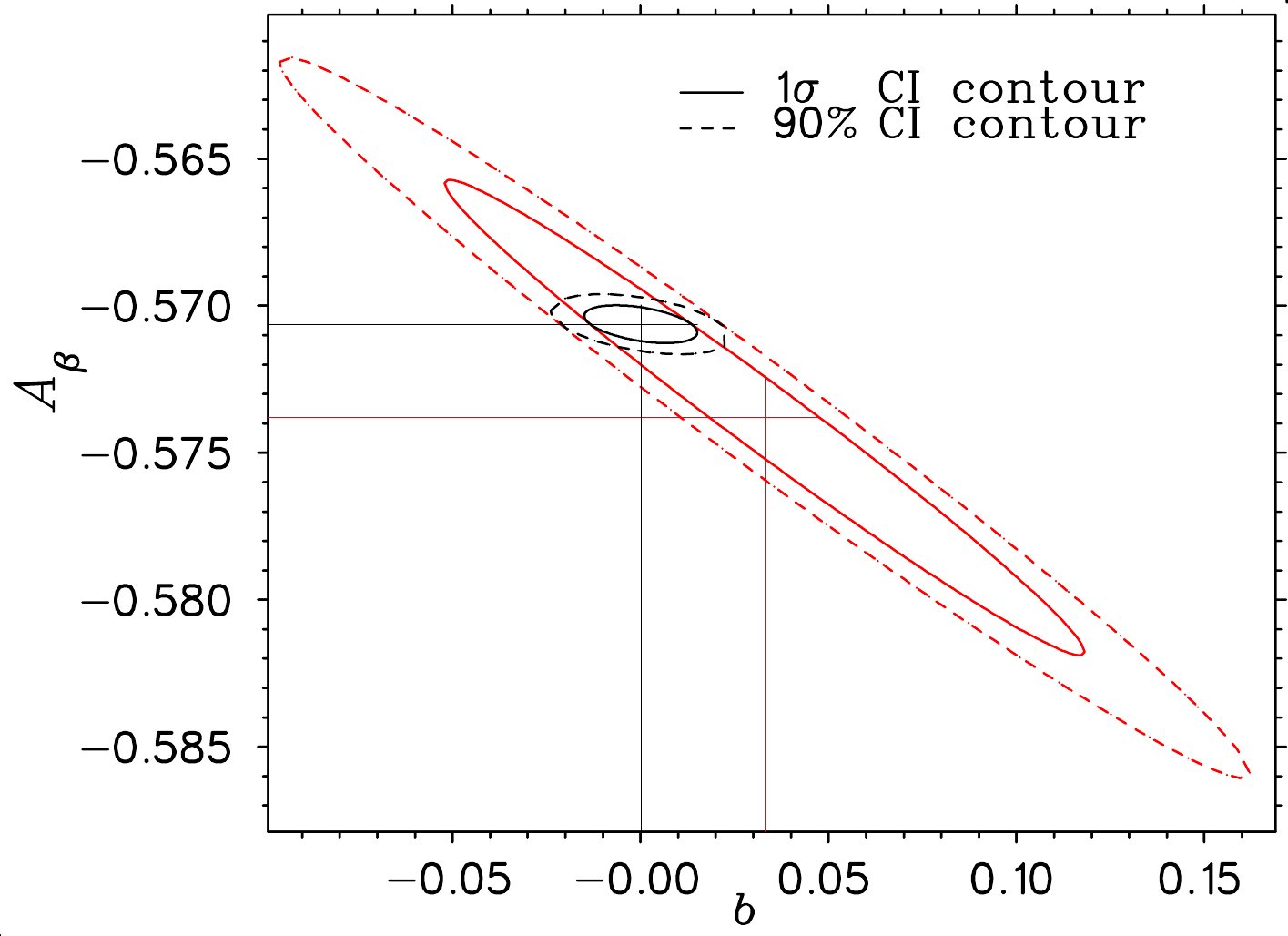}
   \caption{A $\chi^2$ map of all experimental data compared to a simulated parameter space of $A_\beta$ and $b$, showing in solid lines the 1 $\sigma$ 68\% confidence limits from a change in $\chi^2$ of 2.3 from the minimum. Dashed lines are 90\% CL.
       The larger, red ellipses determine the more general measurement of $b$ from Eq.~1.
The tighter, black ellipses add the $chi^2$ contribution from the predicted center value of $A_{\beta}$ from $Ft$ information (see text), which determines $C_S+C'_S$ vs. $C_T+C'_T$ in Fig.~5 below.
         }
    \label{fig:chisq}
\end{figure}

\subsubsection{Check of previous result}



  As a check, by forcing $b$=0, i.e. introducing only the same energy-independent physics allowed by our Ref.~\cite{Fenker2018},
  we deduce 
$A_{\beta}$= -0.5712 $\pm$ 0.013 (stat).
Since forcing $b$=0 produces a 1-parameter fit, we read off the change in $\chi^2$ by 1, not the 2.3 change for the 2-parameter fits.
This deductions implies that our reanalysis in this paper, primarily the time walk experimental correction of Fig.~\ref{fig:timing},
alters our previous answer for the $\beta$ asymmetry with no extra energy dependence from
$A_{\beta} = -0.5707(13)(13)$ to -0.5712(13)(10). The smaller systematic uncertainty is from the near-elimination of the background from untrapped atoms. The centroid change is about one-third of our total uncertainty, so the upper limits on right-handed currents in Ref.~\cite{Fenker2018} and the extraction of $V_{ud}$  are essentially unchanged, and we do not comment further.

\subsection{Systematic uncertainties}

We show systematic uncertainties in Table~I. Most are determined by testing whether a nonzero $b$ is introduced when a parameter is varied in the GEANT4 simulation. We describe a few critical differences from our earlier analysis in Ref.~\cite{Fenker2018}.

{\bf Beta Scattering} The largest systematic uncertainty is from $\beta$ scattering, addressed carefully in our Ref.~\cite{Fenker2018}. Our Supplementary Material showed our own DSSD double-firing events, which provided a solid benchmark to our version of GEANT4 simulation including different low-energy physics lists. This scattering increases at our lower $E_\beta$, potentially introducing false asymmetries, and we find this determines our largest systematic.
We project that by using low-Z materials we can decrease this by an order of magnitude.

{\bf Mirror and DSSD thickness} The uncertainty from mirror thickness and DSSD thickness, which are quite well-determined mechanically, is a cautionary tale. Not all dependence on detector system energy response is cancelled in the asymmetry. We have demonstrated very thin pellicle mirrors in-vacuum and are developing a gas-filled wire chamber for the $\Delta$E detector.

{\bf Low-$E_\beta$ lineshape and energy threshold} The measured asymmetry reflects the standard model's dependence on helicity, the non-unity velocity of the $\beta$ at our lower energies apparent in Fig.~\ref{fig:fit}. So experimental features changing the energy dependence, such as the natural tail in scintillator response from bremsstrahlung and backscatter, introduce additional uncertainty in the energy dependence of the asymmetry. Lowering the $E_{\beta}$ threshold with deomonstrated much thinner mirrors and in-progress $\Delta E$ would also improve our statistical sensitivity to the $m_\beta /E_\beta$ term by a factor of two.

\begin{table}[htb]
  \begin{tabular}{lcc}
    Source & Present & Future\\
           &  $b$             &  $b$ \\
    \hline\\
    $\beta$ Scattering       & 0.031 & 0.003 \\
    Mirror Thickness         & 0.013 & 0.001 \\
    DSSD Thickness           & 0.013 & 0.001 \\
    DSSD Detection Radius    & 0.006 & 0.001 \\
    DSSD signal/noise        & 0.006 & 0.001 \\
    Low-$E_\beta$ lineshape   & 0.008 & 0.008\\
    DSSD XY Energy Agreement & 0.005 & 0.001\\
    DSSD $E$ threshold       & 0.005 & 0.001 \\
    Scintillator Threshold   & 0.004 & 0.001\\
    Scintillator Calibration & 0.003 & 0.003\\
    Atomic Cloud             & 0.002 & 0.002\\
    Background               & 0.004 & 0.004\\
    Be Foil Thickness        & 0.004 & 0.004 \\
    \hline \\
    Total Systematics        & 0.039 & 0.011\\
  \end{tabular}
  \label{fig:errors}
  \caption{Systematic uncertainty budget from the two-parameter analysis for $b$.
    All uncertainties are believed to be uncorrelated, and are added in quadrature. For details on future projections, see text.}
  \end{table}

\section{Results}

Our result for the Fierz interference term is then
$b$ = 0.033 $\pm$ 0.084 (stat) $\pm$ 0.039 (syst).
We plot this 1 $\sigma$ constraint, using Eq.~3 only, on Fig.~5 (the thick black lines).

We also plot (the large black ellipse) our 1 $\sigma$ allowed area considering, as mentioned above, our present work's measured $A_\beta$ and $b$, their correlation through Eqs.~3--4, our collaboration's measured $^{37}$K $Ft$ value~\cite{Shidling2014}, and the average $Ft$ value of $0^+ \rightarrow 0^+$ decays~\cite{TownerHardy2020}. 
This produces a finite constraint on Lorentz tensor couplings and a considerably tighter constraint on Lorentz scalar couplings in Fig.~5.

We explore complementarity of our results to other experiments next.







\subsection{Sensitivity to new physics}

\begin{figure}[htb]
    \centering
    \includegraphics[angle=0,width=1.0\linewidth]{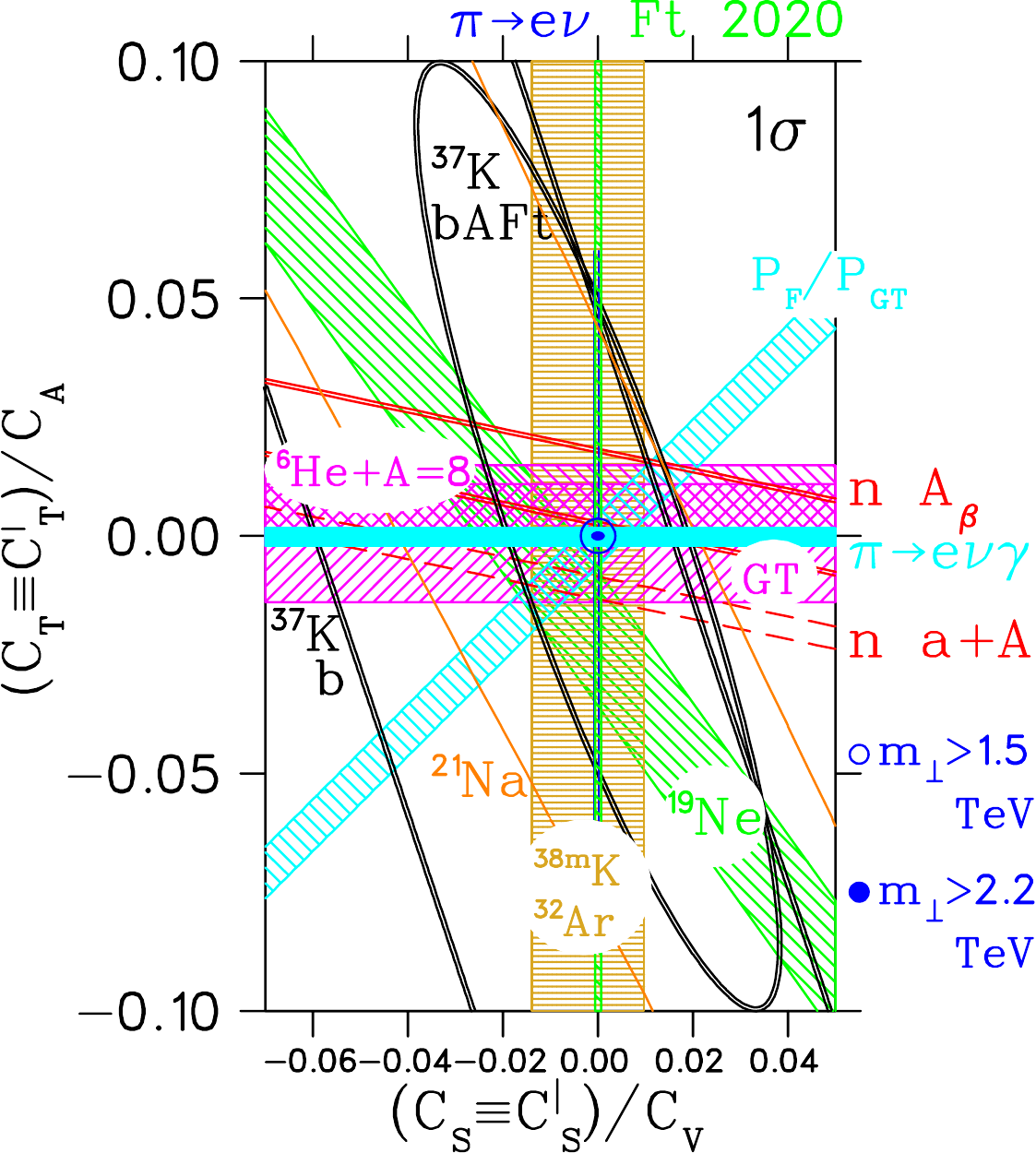}
        \caption{ Exclusion plot at 1 $\sigma$ of Lorentz scalar and tensor lepton-quark couplings to left-handed $\nu$'s and wrong-handed $\beta$'s. Consistency between neutron $\beta$ decay, nuclear $\beta$ decay, and particle physics requires a non-universal interaction with scale less than TeV. Our present $^{37}$K constraints from $b$ (solid black lines), along with our tighter constraints combining $b$, $A_{\beta}$ and $fT$ info (black ellipse), are also detailed in text. 
    }
        \label{fig:scalartensor2}
\end{figure}

We show a
1 $\sigma$
exclusion plot of Lorentz scalar vs. tensor lepton-quark couplings in Fig.~\ref{fig:scalartensor2}, assuming no coupling to non-standard model right-handed neutrinos.
There exist complete reviews setting constraints on the Lee-Yang Lagrangian parameters~\cite{Falkowski2021}, and separately reviews on 2nd-class currents~\cite{Wilkinson2000,Minamisono2011}. To allow for the possibility of non-universal interactions, here we show ten $\beta$ decay and  other particle physics constraints, without review or averaging. Our goal is to show complementarity in situations with less universal physics models than the Lee-Yang Lagrangian.



Our $^{37}$K results could be said to complement the neutron results on $C_S$, because our smaller Gamow-Teller strength weighs $C_S$ more than $C_T$. Yet our own group's direct constraints on $C_S$ from $^{38{\rm m}}$K decay~\cite{BehrGorelov2014} are tighter, shown here averaged with the $^{32}$Ar experiment with similar sensitivity~\cite{Adelberger1999,Garcia2009}.


We show three constraints from particle physics
with computable sensitivity to the Lee-Yang Lagrangian and no sensitivity to 2nd-class currents.
Strong limits are set
on $C_T=C_T'$ from $\pi \rightarrow e \nu \gamma$ (blue cyan band)~\cite{Wauters2014,Bychkov2009}, considerably stronger than the combined limit from $^{6}$He ~\cite{Johnson1963,Gluck1998,GANIL,UW}, $^{8}$B, and $^{8}$Li a$_{\beta \nu}$~\cite{Longfellow2024,Burkey2022}, as well as the combined limit from three Gamow-Teller $A_\beta$ measurements with similar sensitivity due to relatively large $\langle m_\beta/E_\beta \rangle$~\cite{Falkowski2021,Wauters2010,Soti2014,Wauters2009}.

An effective field theory relating $\beta$ decay to an isospin rotation of LHC p+p $\rightarrow$ electron + missing transverse mass $m_{\perp}$ (i.e. a $\nu$)
tightly constrains new physics with $m_{\perp}\geq$ 1.5 TeV from $\sqrt{s}$=8 TeV data (open blue circle)~\cite{GonzalezAlonso2018}, and even more tightly from $\sqrt{s}$=13 TeV data if $m_{\perp}\geq 2.2~{\rm TeV}$~\cite{Falkowski2021,Gupta2018,Aaboud2018,CMS138fb,ATLAS138fb}.

By adding a W in loop diagrams, limits on pseudoscalar interactions from $\pi \rightarrow e \nu $  imply a tight constraint 
$-0.0012 \leq C_S \leq 0.00027$  
at 90\% CL (blue solid vertical lines) for a 200 GeV new boson,
with limits becoming weaker by an order of magnitude for various models (a new boson at the Z mass, scalars coupling purely to right-handed $\nu$'s, or purely imaginary couplings)~\cite{CampbellMaybury2005}.
Both the LHC channel and the $\pi \rightarrow e \nu$ constraints have multiple EFT operators contributing~\cite{Falkowski2021,CampbellMaybury2005} which could in principle cancel in an explicit model of microscopic physics.

The other tight indirect constraints on $C_S$, shown in the green vertical band,
are from the consistency of $0^+ \rightarrow 0^+$ $Ft[E_\beta]$ values in a 2020 global survey~\cite{TownerHardy2020}.
Yet allowing $b$ to float instead of fixing it to zero is known to inflate the uncertainty in $V_{ud}$ by about two~\cite{Falkowski2021}, motivating independent determinations of $b$ with similar sensitivity. Recent theory developments may relax or improve the sensitivity: re-evaluation of the lepton phase space $f$ summarized in Ref.~\cite{SengGorchtein2024} with better calculation of a radiative correction in $^{10}$C decay~\cite{Gennari2025}; continuing theoretical challenges to support existing isospin-breaking calculations, including a possible need for first-principles inclusion of strong interaction isospin breaking~\cite{Konieczka2022} rather than fitting the Nolen-Schiffer anomaly~\cite{TownerHardy2020}.
Note that a fit to Ref.~\cite{TownerHardy2020} of the 2nd-class induced scalar contribution to an electron-nucleon vector current parameterized as
$\frac{e}{A}\frac{m_\beta}{m_{\rm nucleon}} \frac{m_\beta}{E_\beta}$~\cite{Holstein1984,Ormand1989},
produces 
$e = -4 \pm 32$. 

Two experiments are shown for neutron $\beta$ decay. Measurement of $\beta$ decay asymmetry and spectrum shape has competitive sensitivity to $b$ consistent with zero (solid red lines)~\cite{Saul2020}. An analysis including the recoil energy spectrum of protons and Ref.~\cite{Saul2020} data together, but floating $g_A/g_V$ to a different result from either, asserts a 2.8 $\sigma$ nonzero $b$ result (the dashed red line)~\cite{Beck2024}.
  By inspection, the positive result of Ref.~\cite{Beck2024} can be reconciled with the tight constraint on Lorentz tensor lepton-quark interactions from $\pi \rightarrow e \nu \gamma$ decay~\cite{Wauters2014} with a nonzero and negative $C_S$ of -0.05. There are by inspection varying degrees of tension with the above-mentioned Lorentz electron-quark scalar current constraints~\cite{BehrGorelov2014,Adelberger1999,Garcia2009,TownerHardy2020}, with isobaric mirror decays like the present $^{37}$K work, with $^{19}$Ne (from Ref.~\cite{Broussard2014} which used its $fT$ measurement together with $A_{\beta}[E_{\beta}]$~\cite{Calaprice1975}, both under revision~\cite{Combs2020}),  
  and with the relative helicity of $\beta$'s in Fermi vs. Gamow-Teller decay (the cyan band denoted $P_F/P_{GT}$)~\cite{Carnoy1991}.
  Neverthless, we note that interpreting the $C_S \approx$ -0.05 region  as an hypothetical induced scalar for neutron decay with constant form factor $e \approx$ 31 alleviates tension with all other nuclear $\beta$ decay because of the resulting $1/A$ dependence of its contribution. 
 


  We see from Fig.~\ref{fig:scalartensor2} that to be complementary to other physics, we would have to improve our $^{37}$K $b$ result to compete with the best isospin mirror nucleus results in $^{19}$Ne~\cite{Broussard2014,Calaprice1975,Combs2020} and to other measurements.
  As we are presently dominated by statistics, and project to improve systematics, we hope our demonstrated and in-progress improvements in the experiment will let us achieve the order of magnitude needed.


\section{Conclusions}
We have used our $\beta$ asymmetry data from the decay of $^{37}$K to constrain new physics that depends on the $\beta$ energy, measuring the Fierz interference term
 $b$ = 0.033 $\pm$ 0.084 (stat) $\pm$ 0.039 (syst).
We also provide constraints on an exclusion plot on a combination of Lorentz scalar and tensor lepton-quark couplings from our determination of $b$ and $A_\beta$ adding the measured $fT$ of $^{37}$K and its uncertainty~\cite{Shidling2014} (using the center value of $fT$ of $0^+ \rightarrow 0^+$ decays for the Fermi strength).  Our present results using our laser-cooled atoms are dominated by a combination of statistics and solvable $\beta$ scattering, and we hope to improve our statistics-limited result by an order of magnitude to enable complementarity with experiments indicating possible existence of non-standard model physics.



\begin{acknowledgments}

We acknowledge physics input from Mart\'in Gonz\'alez-Alonso.
We acknowledge TRIUMF/ISAC staff, in particular for TiC
target preparation. Supported by the
Natural Sciences and Engineering Research Council of
Canada, 
and the U.S.
Department of Energy, Office of Science, Office of
Nuclear Physics under Awards No. DE-FG03-
93ER40773 and No. DE-FG02-11ER41747.
TRIUMF
receives federal funding via a contribution agreement
through the National Research Council of Canada.
\end{acknowledgments}

\providecommand{\noopsort}[1]{}\providecommand{\singleletter}[1]{#1}%

\end{document}